\documentclass[runningheads]{llncs}

\usepackage[english]{babel}
\usepackage{url}

\usepackage[T2A]{fontenc}
\usepackage[utf8]{inputenc}

\usepackage{amsmath}
\usepackage{amssymb}
\usepackage{mathtools}
\usepackage{subfiles}
\usepackage{graphicx}

\usepackage{graphicx}
\DeclareGraphicsExtensions{.pdf,.png,.jpg,gif}
\usepackage{algorithm}
\let\vec=\mathbf

\newcommand{\vecX}{\mathbf{x}}

\begin{document}

\title{Usage of multiple RTL features for Earthquakes prediction\thanks{Supported by Skoltech}}

\author{P. Proskura\inst{1} \and
A. Zaytsev\inst{2} \and
I. Braslavsky\inst{1} \and
E. Egorov\inst{2} \and
E. Burnaev\inst{2}}
\authorrunning{P. Proskura et al.}

\institute{Insititute of Information Transmission Problems Russian Academy of Science 
\email{polina.231.11@gmail.com}\\
\and
Skolkovo Institute of Science and Technology \\
\email{\{a.zaytsev,e.egorov,e.burnaev\}@skoltech.ru}}

\maketitle 

\begin{abstract}  
We construct a classification model that predicts if an earthquake with the magnitude above a threshold will take place at a given location in a time range $30$-$180$ days from a given moment of time.
A common approach is to use expert forecasts based on features like Region-Time-Length (RTL) characteristics.
The proposed approach uses machine learning on top of multiple RTL features to take into account effects at various scales and to improve prediction accuracy.
For historical data about Japan earthquakes $1992$-$2005$ and predictions at locations given in this database the best model has precision up to $\sim0.95$ and recall up to $\sim0.98$.

\end{abstract}

\section{Introduction}
Nowadays physical modeling fails to provide accurate earthquake predictions because of the complex nonlinear behavior of seismicity. 
Instead researches adopt data-based approaches and construct machine learning or statistics-inspired models based on historical data on earthquakes~\cite{pakistan}.

There are a number of problem statements related to earthquake predictions: consider a target region and predict place and time of the next earthquake, split the region into a grid and predict earthquake at the each of grid nodes, consider each earthquake as a separate event with a given location and time and predict a value of its magnitude~\cite{pakistan}. 
In this paper, we consider the third problem statement and construct a model that predicts if the magnitude exceeds a given threshold for an earthquake at a given location and time. 

A starting point of a model for earthquake prediction is an empirical relationship or a physical model that provides a representation of reality.
Often this representation is not accurate enough, and one adopts a machine learning approach on top of the physics-driven description.

For the earthquakes prediction often several empirical statistical relationships are considered, e.g. Gutenberg–Richter law and Omori-Utsu (O-U) law. 

Gutenberg–Richter law~\cite{b-value} expresses the relationship between the magnitude and a total number of earthquakes of at least that magnitude in a given region:
\[
	\log{N} = a - b M,
\]
where $N$ is the number of events with the magnitude greater than $M$, $a$ and $b$ are commonly referred to as $\vec{b-value}$.
The estimates of coefficients $a$ and $b$ come from historical data.

Omori-Utsu (O-U) law~\cite{O-U-law} represents the decay of an aftershock activity with time  
\[
	\dot{N}(t) = \frac{C_1}{(C_2 + t)^p},
\]
where $t$ is time, $N$ is a number of earthquakes, $C_1$, $C_2$ and decay exponent $p$ (commonly referred to as $\vec{p-value}$) are coefficients fitted using historical data. 
Both these models provide high-level description of earthquakes number in the target region. 
Gutenberg–Richter law provide a number of earthquake in a region.
Omori-Utsu law provides a connection between the past and the future seismologic activity in a region.

Another physics-inspired description is RTL features~\cite{RTL-Sobolev} that provide aggregation of the past seismic activity into a single feature by weighting past earthquakes that occur near the region
where we want to predict an earthquake.
RTL features also have a number of hyperparameters to be fitted using historical data.

On top of these features we can construct a machine learning model~\cite{panakkat2007neural,rouet2017machine,asim2017earthquake,asencio2016sensitivity}.
An example of such work~\cite{asim2017earthquake} considered the prediction of earthquakes as a binary classification problem. 
Authors generated $51$ meaningful seismic features based on well-known seismic characteristics such as ``Standard Deviation of b-value'' or ``Time ($T$) of $n$ events''. As models they used various ensemble methods such as Random Forest, Rotation Forest and RotBoost. 

In this paper we consider a different problem statement: we want to predict earthquakes at a given location in a given time interval.

As input features we use normalized RTL features with different parameters and at different time scales. We examine a number of machine learning techniques to make use of generated features.

When developing models we take into account peculiarities of the problem: imbalance of classes (there are only a few large earthquakes in the dataset)~\cite{burnaev2015model,burnaev2015influence,Imbalance2019}; data is represented in time-series format; no external features are given and we need to generate features from the data on past earthquakes.

\section{Problem statement}

We consider a historical data about Japan earthquakes in $1992$-$2005$. In the given dataset each earthquake has four parameters: location $(x, y)$, time $t$ and magnitude $M$.
Let us denote by $ c(x, y, t)$ an indicator function that the earthquake $e$ takes place at some location $(x, y, t)$. We set $c(x, y, t) = 1$ if there is at least one earthquake of magnitude $M_e \geq M_c$ for some threshold $M_c$ with coordinates $(x_e, y_e, t_e)$ satisfying the following constraints:
\begin{equation}
\label{eq:target_defintion}
\|(x, y) - (x_e,y_e)\|_2 	\leq R_c,~~
\delta_c < t_e - t < T_c.
\end{equation}
Otherwise we set $c(x, y, t) = 0$.

Our goal is to construct a model that predicts if there is an earthquake in the time cylinder $[T + T_{\min}, T + T_{\max}]$ for some $T$ using historical information about all earthquakes up to time $T$.

\begin{figure}
    \centering
    \includegraphics[scale=0.17]{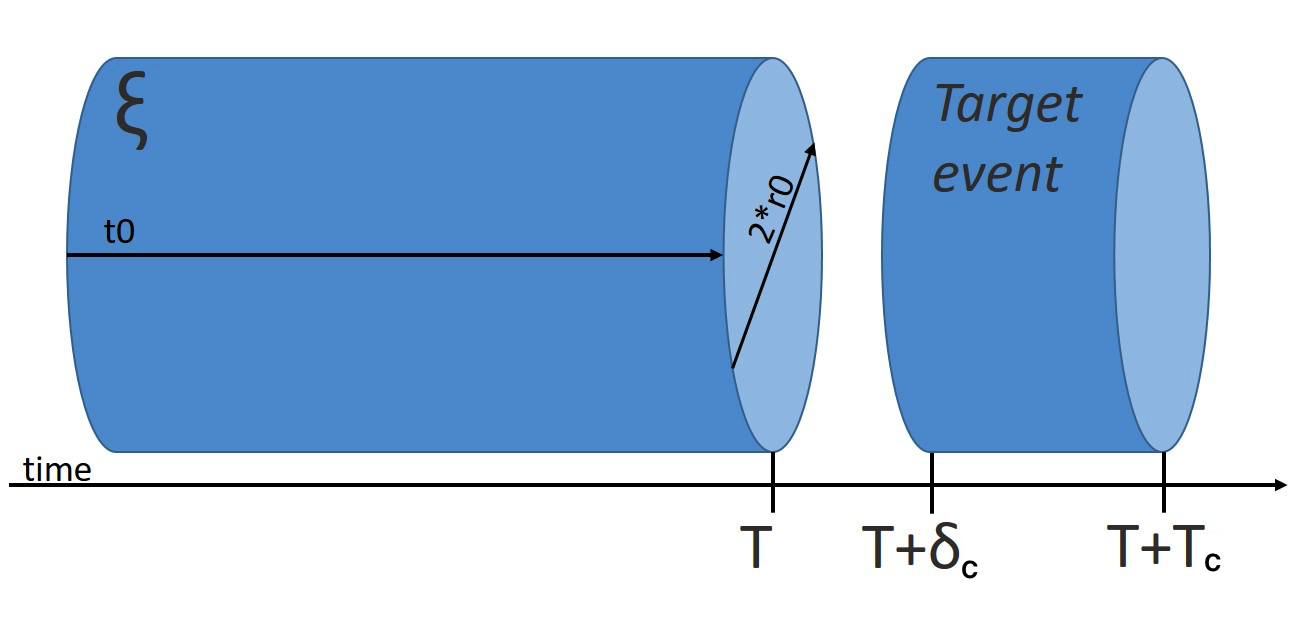}
    \caption{Space-time cylinder used for generation of RTL features and interval considered for prediction}
    \label{fig:cylinder_RTL}
\end{figure}

We aggregate information about earthquakes up to time $T$ in a vector of features $\vecX$ of a fixed length. 
In particular we generate RTL features using procedure described in Subsection~\ref{sec:rtl_features}.
Figure~\ref{fig:cylinder_RTL} depicts 
a space-time cylinder used for the generation of input features and the target interval for prediction.

Finally for all earthquakes in the database we get pairs $(\vecX_i, c_i)$, where $c_i$ is defined by \eqref{eq:target_defintion} with the magnitude threshold $M_c = 5$.
All these pairs form a sample $D = \{(\vecX_i, c_i)\}_{i = 1}^n$ of size $n$.
Our goal is to create a model $\hat{c}(x, y, t)$ approximating the actual $c(x, y, t)$.



\section{Data}

\begin{figure}
    \centering
    \includegraphics[scale=0.5]{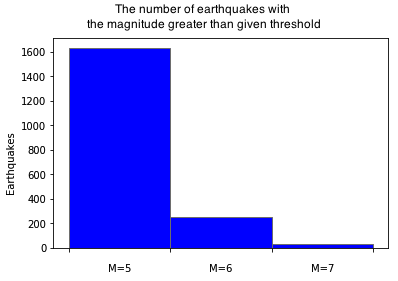}
    \caption{Histogram of magnitudes: the total sample size is about $300 000$; there are only about $2000$ earthhquakes with a magnitude greater or equal to $5$.}
    \label{fig:hist_magnitudes}
\end{figure}

We study the prediction of strong earthquakes in the middle-term horizon. Strong earthquake is an earthquake with the magnitude higher than $M_c = 5$. Prediction of earthquakes is difficult not only because it is a very complex non-linear process, but even the historical dataset has challenging structure:
\begin{itemize}
\item In Japan from $1990$ to $2016$ there were $247,204$ earthquakes, however the sample is very unbalanced: see Figure~\ref{fig:hist_magnitudes} with a distribution of earthquake magnitudes. Most of classifiers and their accuracy metrics are tailored to balanced samples with uniform distribution of examples among classes. Thus in our case we have to tune a classifier to make it more sensitive to the target class,
\item The dataset is non-homogeneous as the
network of seismic stations changes over time and is nonuniform. So we should take these into account when generating features and assessing the results provided by predictive models.
\end{itemize}

\section{Methods}

\subsection{RTL features}
\label{sec:rtl_features}
As inputs for Machine Learning model we use RTL features. 
The basic assumption of the $\vec{Region-Time-Length}$ $\vec{(RTL)}$ algorithm~\cite{RTL-Sobolev} is that the influence weight of each prior event on the main event under investigation may be quantified in the form of a weight. Weights become larger when an earthquake is larger in magnitude or is closer to the considered location or time. Thus, $\vec{RTL}$ characterizes the level of seismicity at a point of space in a certain time. 

The $\vec{RTL}$ takes into account weighted quantities associated with three parameters (time, place and magnitude) of earthquakes. 
An $\vec{RTL}$ parameter is defined as the product of the following three functions:
\[
	\label{RTL Precursor}
	\mathsf{RTL}(x, y, t, M) = \mathsf{R}(x, y, t, M) \cdot \mathsf{T}(x, y, t, M) \cdot \mathsf{L}(x, y, t, M),
\]
where $\mathsf{R}(x, y, t, M)$ is an epicentral distance, $\mathsf{T}(x, y, t, M)$ is a time distance and 
$\mathsf{L}(x, y, t, M)$ is a rupture length. 
They depend on the size of the space-time cylinder $\mathcal{E}_{r_0, t_0}$, defined by radius $r_0$ and time length $t_0$, see Figure \ref{fig:cylinder_RTL}:

\begin{align*}
\mathsf{R}(x, y, t, M) &= \sum\limits_{e_i \in \mathcal{E}}\exp\left(-\dfrac{r_i}{r_0}\right), \\
\mathsf{T}(x, y, t, M) &= \sum\limits_{e_i \in \mathcal{E}}\exp\left(-\dfrac{t - t_i}{t_0}\right), \\
\mathsf{L}(x, y, t, M) &= \sum\limits_{e_i \in \mathcal{E}}\left(\dfrac{l_i}{r_i}\right),
\end{align*}
where $e_i$ is a full description of an earthquake~$(x_i, y_i, t_i, M_i)$, $l_i$ is an empirical relationship specific for Japan \[\log l_i = 0.5 M_i - 1.8,\]  and $r_i = \sqrt{(x - x_i)^2 + (y - y_i)^2}$.  We consider only earthquakes with magnitude at least $M_i \geq M_c = 5$.

\subsection{Normalization of RTL features}
$\vec{RTL}$ is a very unstable statistics. Therefore in \cite{RTL-Huang} they proposed to normalize the parameters. 

We perform normalization during data preprocessing stage. 
For our case we transform the data to make each feature zero-mean and unit-variance~\cite{Norm}:
\begin{itemize}
    \item Calculate mean and standard variance for each feature,
    \item Subtract the mean from each feature,
    \item Divide the values of each feature by its standard deviation.
\end{itemize}

Another option is to subtract a moving average instead of the mean value. It helps to take into account a trend in time. 

\begin{figure}
    \centering
    \includegraphics[scale=1.0]{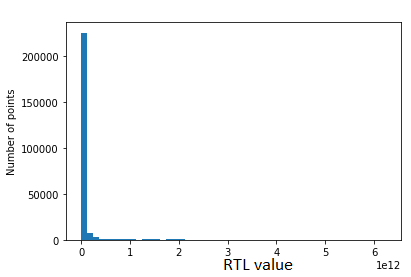}
    \caption{Histogram of RTL values for each point. Small values prevail}
    \label{fig:rtl_hist}
\end{figure}

The negative $\vec{RTL}$ means a lower seismicity compared to the background rate around the investigated place, and the positive $\vec{RTL}$ represents a higher seismicity compared to the background. See the histogram of TRL values in Figure ~\ref{fig:rtl_hist}. We are interested in both types of anomalies.      

\subsection{Classifiers}
As classifiers we use the following machine learning methods:
\begin{itemize}
\item $\vec{Major~RTL}$ is the method which estimates a threshold for RTL features. We estimate the optimal threshold from the train sample. If the value of a feature is less than that threshold, we set the label to $0$, and to $1$ otherwise. Using the estimated threshold we can classify new data points. 
\item $\vec{Logistic~regression}$ is a statistical model used to predict the probability of an event occurrence. The model has the form:
\[
	Pr(c = 1 | \vecX ) = f(\boldsymbol{\theta}^{\top} \vecX + b),
\]
where $f(z) = \frac{1}{1+e^{-z}}$ is a logistic function, and $\boldsymbol{\theta}$ and $b$ are the parameters of the model estimated from data~\cite{logregr}.
\item $\vec{Random~Forest}$ is an ensemble classifier that is developed on the basis of majority voting of decision trees. Various number of decision trees are generated over bootstrap
samples of the training dataset. The final decision is based on aggregation of the predictions obtained by all the decision trees. Thus, a Random Forest allows to find complicated relationships in the data, but at the same time more resistant to overfitting ~\cite{randfor}. 
\item $\vec{AdaBoost}$ is another method that combines classifiers into ensembles. The key feature of the method is that it introduces weights for all objects. At each iteration the weights of each incorrectly classified object increase, so a new classifier ensemble focuses attention on these objects. AdaBoost is sensitive to noise and it is less prone to overfitting compared to other algorithms of machine learning~\cite{adaboost}.
\item $\vec{Gradient~Boosting}$ is an ensemble method that builds an ensemble of trees one-by-one, then the predictions of the individual trees are aggregated. The next decision tree tries to approximate the discrepancy between the target function and the current ensemble prediction by
reconstructing the residuals. Thus, an iterative process is used; in each iteration the loss function is minimized by the gradient descent ~\cite{gradboost}.
A nice feature of Gradient Boosting is ability to treat imbalanced-classification problems after a simple modification~\cite{kozlovskaia2017deep}.
\end{itemize}

\subsection{Resampling techniques}
\label{ssssss}

The problem at hand is imbalanced: the number of earthquakes with big enough magnitudes is small.
So it is natural to apply machine learning heuristics that can deal with a natural class imbalance, see \cite{burnaev2015influence,Imbalance2019}.
Here we consider two simple yet efficient techniques for resampling: we modify our initial training sample to make more emphasis on minor class objects
\begin{itemize}
    \item Oversampling --- increase weights of the minor class objects in a random way,
    \item Undersampling --- drop some major class objects to balance number of instances of each class in the training sample.
\end{itemize}

\section{Results}

Selection of used classification accuracy metrics is motivate by imbalanced nature of the considered problem. 
In addition to commonly used Precision, Recall and ROC AUC we use F1-score and Precision-Recall AUC (PR AUC).
See definitions of the accuracy metrics in Appendix~\ref{sec:quality_metrics}.

In the experiments below we use  $M_c = 5$, $R_c = 50$ (km), $\delta_c = 10$ (days), $T_c = 180$ (days).

\subsection{Models for RTL features generated with a single pair of hyperparameters $(r_0, t_0)$}
\label{subsub}

We use the following  sets of hyperparameters to calculate $\vec{RTL}$ features:
\begin{equation}
\vec{r}_0=[10, 25, 50, 100],~ 
\vec{t}_0 = [30, 90, 180, 365].
\label{eqeq}
\end{equation}
So we generate $4 \times 4 = 16$ different types of features for different values of $(r_0, t_0)$.
For each pair $(r_0, t_0)$ we generate $20$ different autoregression features: we consider not only the current time moment $t$, but also the moments $t - 1, t - 2, \ldots, t - 19$. We use these $20$ features as input for our machine learning model.

In Table~\ref{table:results} we provide results for all considered values of $r_0$ and some values of $t_0$. In the table we include results for those values of $t_0$ given fixed $r_0$, which maximizes ROC AUC of the model among all $t_0$ values, listed in \eqref{eqeq}.

We see that features with sufficiently big $r_0$ values and sufficiently small $t_0$ values provide the best accuracy.

The best model according to our experiments is obtained via Gradient Boosting ~\cite{xgb}.
Gradient Boosting works better than linear Logistic Regression because the magnitude of earthquakes depends on RTL features in a non-linear way. 
Both these approaches work better than Major RTL, i.e. more sophisticated machine learning approaches work better than a simple threshold rule for the problem of earthquakes prediction.


\begin{table}
\centering
    \begin{tabular}{lllllllll}
    \hline
    $r_0$ & best $t_0$ & Algorithm & Precision & Recall & F1 & ROC AUC & PR AUC\\ \hline
    10 & 180 & Logistic Regression & 0.54 & 0.36 & 0.43 & 0.64 & 0.53\\
    & & Random Forest & 0.62 & 0.51 & 0.56 & 0.76 & 0.69\\
    & & AdaBoost & 0.63 & 0.50 & 0.56 & 0.77 & 0.83\\
    & & Gradient Boosting  & 0.62 & 0.52 & 0.56 & 0.80 & 0.69\\
    & & Major\_RTL & 0.57 & 0.47 & 0.47 & 0.52 & 0.77 \\ \hline
    25 & 90 & Logistic Regression & 0.72 & 0.58 & 0.64 & 0.70 & 0.44\\
    & & Random Forest & 0.79 & 0.51 & 0.62 & 0.74 & 0.68\\
    & & AdaBoost & 0.75 & 0.67 & 0.71 & 0.77 & 0.53\\
    & & Gradient Boosting  & 0.82 & 0.6 & 0.69 & 0.79 & 0.77\\
    & & Major\_RTL & 0.67 & 0.52 & 0.62 & 0.70 & 0.60 \\ \hline
    50 & 180 & Logistic Regression & 0.91 & 0.84 & 0.88 & 0.83 & 0.67\\ 
    & & Random Forest & 0.91 & 0.84 & 0.87 & 0.80 & 0.70\\
    & & AdaBoost & 0.83 & 0.96 & 0.89 & 0.81 & 0.74\\
    & & Gradient Boosting  & 0.89 & 0.92 & 0.91 & 0.87 & 0.73\\
    & & Major\_RTL & 0.60 & 0.74 & 0.74 & 0.80 & 0.71 \\ \hline
    100 & 180 & Logistic Regression & 0.94 & 0.94 & 0.94 & 0.80 & \textbf{0.94}\\
    & & Random Forest & \textbf{0.97} & 0.90 & 0.94 & 0.89 & 0.90\\ 
    & & AdaBoost & 0.96 & 0.96 & 0.96 & 0.92 & 0.90\\
    & & Gradient Boosting & 0.95 & \textbf{0.98} & \textbf{0.97} & \textbf{0.93} & \textbf{0.94}\\
    & & Major\_RTL & 0.90 & 0.86 & 0.89 & 0.83 & 0.89 \\ \hline
    \end{tabular}
\caption{Results for different values of hyperparameters of generated RTL features: we get better results are for a bigger size of the cylinder}
\label{table:results}
\end{table}

\subsection{Aggregation of RTL features}
In this section we used $16$ calculated RTL features as inputs for machine learning models. In this case the input dimension is equal to $20 \times 16 = 320$.
We provide accuracy of obtained models in Table ~\ref{table:set_results}. 
The improvement compared to the single best RTL feature  is insignificant (see results in subsection \ref{subsub} above).
We conclude that it is not possible to improve accuracy by using aggregation of RTL features.

\begin{table}
\centering
    \begin{tabular}{lllllll}
    \hline
    Algorithm & Precision & Recall & F1 & ROC AUC & PR AUC\\ \hline
    Gradient Boosting (best single RTL) & 0.97 & 0.96 & 0.96 & 0.95 & 0.94 \\
    Major\_RTL & 0.90 & 0.85 & 0.89 & 0.82 & 0.89 \\ 
    Logistic Regression & 0.94 & 0.95 & 0.94 & 0.81 & 0.94\\
    Random Forest & 0.97 & 0.90 & 0.94 & 0.89 & 0.90\\
    AdaBoost & 0.97 & 0.96 & 0.97 & 0.92 & 0.90\\
    Gradient Boosting & 0.95 & 0.98 & 0.97 & 0.93 & 0.93\\
\hline
    \end{tabular}
    \caption{Results for multiple RTL features generated with different hyperparameters}
    \label{table:set_results}
\end{table}

\subsection{Usage of resampling techniques}

There are a number of resampling techniques that can deal with imbalanced classification problems~\cite{burnaev2015model,burnaev2015influence,Imbalance2019}, see subsection \ref{ssssss} above.
Here we consider classification with oversampling and undersampling, as well as
no-resampling case.
For each case we provide results in  Table~\ref{table:imb_results} for the  machine learning model with the best overall performance.

We see that both undersampling and oversampling improve the quality of the models.

\begin{table}
\centering
    \begin{tabular}{llllll}
    \hline
    Approach & Algorithm & Precision & Recall & F1 \\ \hline
    No resampling      & Gradient Boosting & 0.84 & 0.59 & 0.59  \\
    Oversampling  & Random Forest & 0.94 & 0.99 & 0.97  \\
    Undersampling & Random Forest & 0.96 & 0.87 & 0.91\\
  \hline
    \end{tabular}
    \caption{Model quality when using imbalanced classification based on resampling. In these experiments we used a training sample of a smaller size}
    \label{table:imb_results}
\end{table}

\section{Conclusion}

We considered the problem of middle-term earthquakes prediction.
Usage of Machine learning provides an improvement compared to the state-of-the-art {major RTL} method. In particular the model based on Gradient Boosting with RTL features as inputs delivers the best performance. It is interesting that for many cases RTL features, generated using a single set of  hyperparameters, provide results not worse than results for multiple RTL features, generated using multiple sets of hyperparameters. Also we demonstrated that imbalanced classification approaches can improve accuracy.

We can further improve accuracy by considering the following avenues for future research:
\begin{itemize}
\item earthquakes with big magnitude are rare events, a kind of anomalies. Thus we can first detect sequences of anomalies of different types in the historical stream of earthquake data \cite{Degradation2016,ConformalAD2015,Multichannel2017,kNN2017,ConformalMartingales2017}, and then we can construct ensembles for rare events prediction \cite{EnsemblesDetectors2015,newsmolyakov} using detected anomalies and their features as precursors of major earthquakes to optimize specific detection metrics similar to the one used in \cite{Vehicle2017},
    \item use privileged information about the future events, which is accessible during the training stage. Analogous approach, used in \cite{OCSVM2016,OCSVM2018} for anomaly detection, allowed significant accuracy improvement,
    \item historical data on earthquakes has a spatial component, thus a graph of dependency between streams of events, registered by different ground stations can be constructed and modern methods for graph feature learning \cite{ICML} and panel time-series feature extraction \cite{TDA,MF2018} can be applied to enrich the set of input features, used for predictive model construction.
\end{itemize}

\section*{Acknowledgements}
The research was partially supported by the Russian Foundation for Basic Research grant 16-29-09649 ofi m.

\bibliographystyle{splncs04}
\bibliography{main}

\appendix
\section{Quality metrics for classification problem}
\label{sec:quality_metrics}

Let us introduce necessary definitions. Classification problem can be formulated as whether the object belongs to the target class or not.  There are four types of classification errors:
\begin{itemize}
    \item True Positive --- if the object belongs to the target class and we predict that it belongs,
    \item True Negative --- if the object does not belong to the target class and we predict that it does not,
    \item False Positive --- if the object does not belong to the target class but we predict that it does,
    \item False Negative --- if the object belongs to the target class but we predict that it does not.
\end{itemize}

The \textbf{precision} score quantifies the ability of a classifier to not label a negative example as positive. The \textbf{precision} is the
probability that a positive prediction made by the classifier is positive. The score is in the range $[0,1]$ with $0$ is the worst, and $1$ is perfect. 
The precision score can be defined as:
$$\textbf{Precision} = \frac{\mbox{True Positive}}{\mbox{True Positive} + \mbox{False Positive}}$$ 

The \textbf{recall} score quantifies the ability of the classifier to find all the positive samples. It defines what part of positive samples have been chosen by classifier as positive. The score is in the range $[0,1]$ with $0$ is the worst, and $1$ is perfect. 
$$\textbf{Recall} = \frac{\mbox{True Positive}}{\mbox{True Positive} + \mbox{False Negative}}$$ 

The \textbf{F1}-\textbf{score} is a single metric that combines both precision and recall via their harmonic mean. It measures the test accuracy and reaches its best value at $1$ (perfect precision and recall) and worst at $0$:
$$\textbf{F1} = 2 \frac{\mbox{Precision}\cdot \mbox{Recall}}{\mbox{Precision} + \mbox{Recall}}$$

\textbf{ROC AUC score} counts the area under the ROC-curve. ROC-curve is the curve, which shows how True Positive rate depends on the False Positive rate. These characteristics are defined as 
$$\mbox{True Positive Rate} = \frac{\mbox{True Positive}}{\mbox{True Positive} + \mbox{False Negative}},$$ 
$$\mbox{False Positive Rate} = \frac{\mbox{False Positive}}{\mbox{False Positive} + \mbox{True Negative}}.$$ 

ROC AUC score measures the quality of binary classifier. The best value is $1$, value $0.5$ corresponds to a random classification. 

\textbf{PR AUC score} counts the curve area under the Precision-Recall curve, which characterizes how Precision depends on Recall. Precision-Recall is a useful classification measure when the classes are imbalanced. The perfect classifier curve ends in $(1.0, 1.0)$ and has the area under it that is equal to $1$.

\end{document}